# 3. Does the risk of death continue to rise among supercentenarians?

Francisco Villavicencio*[1] and José Manuel Aburto[1]

## Abstract

A recent study in the *2017 Living to 100 Monograph* published by the Society of Actuaries suggests, in contrast to previous research, that the risk of death after 110 increases with age. By fitting a Gompertz model to estimated central death rates for the oldest old, the authors challenge existing theory and empirical research indicating a deceleration of mortality at older ages and the emergence of a plateau. We argue that their results are inconclusive for three reasons: (1) the data selection was arbitrary; (2) the statistical analysis was inappropriate; and (3) the presentation of the results is misleading and inadequate. We therefore claim that the hypothesis that the human force of mortality increases after age 110 has not been proved.

* Correspondence: fvillavicencio@health.sdu.dk
[1] Interdisciplinary Center on Population Dynamics, University of Southern Denmark, Odense, Denmark.

## 3.1 Introduction

Does the human force of mortality increase after age 110? A recent study by Gavrilova et al. (2017) suggests, in contrast to previous research, that this may effectively be the case. By fitting a Gompertz model to estimated central death rates for the oldest old, they aim to prove that these rates still increase with age, and challenge existing theory and empirical research indicating a deceleration of mortality at older ages and the emergence of a plateau (Gampe 2010, Robine and Vaupel 2001, 2002). Despite the efforts made by Gavrilova and colleagues to validate their hypothesis, we believe that their results are inconclusive for three reasons that we discuss in the following: (1) the data selection was arbitrary; (2) the statistical analysis was inappropriate; and (3) the presentation of the results is misleading and inadequate. The main flaw in this study is that the authors focused on the analysis of cohorts with no survivors beyond age 115, and systematically assumed that the probability of death at that age is 1. Furthermore, they do not mention any considerations about the uncertainty of the estimated central death rates.

We have carried out our analyses using the open-source statistical software R (R Core Team 2017). The results and the figures presented here are fully reproducible from the code and data available in the supplementary materials.

## 3.2 Selection of the data

Gavrilova et al. (2017) analyzed data from two sources: the International Database on Longevity (IDL 2017) and the Gerontology Research Group Database on Supercentenarians (GRG 2017). The IDL contains all of the validated records of individuals aged 110 years and older – the so-called supercentenarians – from 15 countries, such that the inclusion of a person in the database does not depend on his or her age. The GRG aims to authenticate cases of the oldest humans in history, but its data may not be suitable for analyzing age patterns of mortality. For instance, the probability of being considered for inclusion in the GRG database increases with age, as older people get more attention in the media. Thus, individuals who died at ages 110 or 111 may be underrepresented in the GRG. Accordingly, our analysis of the work by Gavrilova and colleagues is restricted to the results they obtained with data from the IDL.

The data that were publicly available from the IDL as of 29 November 2017 were last updated on June 2010, and the last observed death dates from 2007 (IDL 2017). This dataset includes 672 supercentenarians born between 1852 and 1898. We have made these data available in the supplementary materials because they correspond to those used by Gavrilova et al. (2017). To avoid dealing with censored individuals, the only data Gavrilova and colleagues considered in their analyses were from cohorts born in 1894 or earlier, whom they believed to be extinct. It is our view, however, that this assumption is inaccurate, because one of the supercentenarians who were reported alive in the IDL from Germany was born in 1894. Thus, it does not appear to be the case that all of the cohorts in the database born in 1894 or earlier were extinct at the time of data collection. It is unclear how the authors dealt with this individual; that is, whether they assigned her a date of death extracted from another source, or simply excluded her from the study. In addition, their analytical strategy did not account for the presence of right-truncated individuals due to country-specific sample designs – which, as Gampe (2010) has pointed out, could affect mortality estimates.

These problems notwithstanding, let us assume for the sake of convenience that we are dealing exclusively with cohorts who are extinct. Gavrilova and colleagues went one step further and divided the data into cohorts born in 1852-1884 and cohorts born in 1884-1894, while focusing on the latter. They justify this decision by arguing that cohorts born in 1884-1894 "have the largest number of cases in IDL (401) and hence are likely to be more complete" (Gavrilova et al. 2017, p. 4). However, this division of the data is arbitrary, and seems to hide a certain degree of intentionality, as the 400 dead supercentenarians from the IDL who were born in 1884-1894 (excluding the living individual from Germany) died between the ages of 110 and 115. If Gavrilova and colleagues had extended the analysis to the cohorts 1880-1894, they would have had to include 115 additional individuals, one of whom died at the age of 119. Moreover, if they had extended the interval to 1875-1894, they would have ended up with a total of 600 supercentenarians, among whom are three individuals who died at ages 117, 119, and 122, respectively (IDL 2017).

Selecting a subset of the data that only includes supercentenarians who died at age 115 or younger, and ignoring those individuals who survived beyond that age, is an example of selection bias. Such a bias may have a strong effect on the analysis, leading to incorrect results and compromising the validity of the conclusions.

## 3.3 Statistical methods

Gavrilova et al. (2017) claim to have proved wrong that human mortality after age 110 is flat, a hypothesis they attribute to Gampe (2010). This is, however, a misinterpretation of Gampe's results, since in her conclusion she states that death rates are constant between the ages of 110 and 114 only, and that beyond age 114 the data become too sparse to make reliable statements (Gampe 2010). Moreover, they complain that "Gampe wrote her own program for hazard rate calculation, rather than using estimates provided by standard statistical packages, so it is difficult to test and reproduce her results" (Gavrilova et al. 2017, p. 15). We believe this criticism is unjustified for two main reasons. First, Gampe (2010) provides a complete formal mathematical description of her model, including details on the likelihood function and the different types of data (sampling frames of truncated and censored observations). This information can be used to implement the model with the preferred statistical software. Second – and perhaps more importantly – because built-in software packages are sometimes not flexible enough for the analysis of data with certain particularities, and it is essential in such situations to proceed with care. This is the case for the age interval death rates computed by Gavrilova and colleagues using existing functions from a commercial statistical software package. Letting $D_x$ denote the deaths within the age interval $[x, x + \Delta x]$, and $N_x$ the number alive at the beginning of that interval, the corresponding central death rates are estimated as

$$m_x = \frac{1}{\Delta x} \frac{q_x}{1 - q_x/2} , \qquad (1)$$

where $\Delta x$ is the length of the age interval and $q_x = D_x/N_x$ (Gavrilova et al. 2017, p. 4, although a slightly different notation is used here).

This method for assessing death rates provides a reasonable estimation for large samples, but turns out to be inappropriate in the analysis of the data on supercentenarians. First, the method implies that the probability of death in the last age is always $q_x = 1$, and the corresponding central death rate $m_x = 2$ when $\Delta x = 1$. This is wrong because the IDL reports validated individuals who lived up to age 122 (Jeanne L. Calment, 1875-1997), and these individuals are ignored by restricting the analysis to cohorts who died at age 115 or younger (1884-1894). While reaching higher ages may be unlikely, building a model that assumes that the probability of death at age 115 is 1 is equivalent to the assumption that no human being can survive beyond that age, which is false. In addition, as we will discuss in greater detail later, it is important to note that the estimated central death rates for the highest ages are not trustworthy due to the scarcity of data. For instance, only three individuals from the 1884-1894 birth cohorts reached age 115 (IDL 2017), and attempting to compute a rate with only three observations is highly questionable. Accepting the limitations of the data, Gampe (2010) concluded that her results are reliable for ages 110 to 114 only. Gavrilova et al. (2017), by contrast, did not mention any such considerations.

As a final remark, note that the IDL provides data on a daily time scale – meaning that it is possible to know how many days a supercentenarian lived after his or her last birthday – and it is worth using that information. Gavrilova and colleagues provide an estimation of central death rates for the 1884-1894 birth cohorts from the IDL in quarter-year age intervals (Gavrilova et al. 2017, Fig. 4), but most of their analysis focused on single-year estimates.

## 3.4 Analysis of the results

When publishing a paper, providing replicable results is always good practice. When criticizing the replicability of someone else's work, it is a must. The results presented by Gavrilova et al. (2017) are confusing and misleading, and we have been able to reproduce only some of their findings after making additional guesses not detailed in their manuscript.

### 3.4.1 Estimates of the Gompertz parameters

We begin by looking at Table 1, which reproduces Table 2 in Gavrilova et al. (2017). This table shows their estimates for the parameters of the Gompertz model, fitted to the central death rates of five subgroups of supercentenarians from the IDL: birth cohorts 1884-1894; cohorts 1884-1894 born in the USA; birth cohorts 1884-1894 with high-quality age validation; birth cohorts 1884-1894 measured in quarter-year age intervals; and cohorts born before 1885.

| Subgroup | Slope parameter | Intercept parameter |
|---|---|---|
| Birth cohorts 1884-1894 | | |
| All | 0.163 <br> $(0.047, 0.279)$ | $9.61 \times 10^{-9}$ <br> $(-1.15 \times 10^{-7}, 1.34 \times 10^{-7})$ |
| Born in the USA | 0.204 <br> $(0.071, 0.337)$ | $9.76 \times 10^{-11}$ <br> $(-1.35 \times 10^{-9}, 1.54 \times 10^{-9})$ |
| All in group A (high quality data) | 0.165 <br> $(0.043, 0.287)$ | $8.03 \times 10^{-9}$ <br> $(-1.01 \times 10^{-7}, 1.17 \times 10^{-7})$ |
| All, quarter-year age intervals | 0.214 <br> $(0.073, 0.355)$ | $3.22 \times 10^{-11}$ <br> $(-4.76 \times 10^{-10}, 5.40 \times 10^{-10})$ |
| Older birth cohorts born before 1885 | | |
| All | 0.018 <br> $(-0.072, 0.108)$ | 0.095 <br> $(-0.853, 1.043)$ |

**Table 1.** Reproduction of Table 2 in Gavrilova et al. (2017): Parameters of the Gompertz model fitted to five subgroups of supercentenarians from the IDL (2017). Values between parentheses represent 95% confidence intervals.

In Table 1, the Gompertz parameters are labelled "slope" and "intercept", respectively. Nevertheless, rather than providing estimates of the intercept parameter, Gavrilova et al. (2017) provide its exponential. The force of mortality (hazard rate or risk of death) of the Gompertz model is usually expressed as

$$\mu(x) = e^{a+bx} = Ae^{bx}, \qquad (2)$$

where $x$ corresponds to age, and $a$, $b$, and $A$ are parameters, with $A = e^a$. In a natural logarithmic scale, the Gompertz force of mortality becomes a linear equation in which one intuitively identifies $b$ as the slope and $a$ as the intercept. However, Gavrilova et al. (2017) provide values of $A$ instead of $a$. This is confusing, and has a strong effect on how we read their results. In the estimates of parameter $A$ (right column in Table 1), the confidence intervals include negative values. We should therefore contemplate the possibility that $A < 0$. But following (2), if $A$ is negative, so is the force of mortality, which contradicts the definition of death rate. While trying to reproduce their results, we have not been able to recover these same confidence intervals.

Following (1), Gavrilova et al. (2017) estimated central death rates for each of the five subgroups of supercentenarians described in Table 1. Next, they estimated parameters $A$ and $b$ by fitting the Gompertz model defined in (2) to each of these subgroups with a weighted non-linear regression. We have reproduced this procedure in R (R Core Team 2017) and using the same software as Gavrilova et al. (2017), obtaining identical results in both cases (the R code is available in the supplementary materials). Still, there is not an exact match between our estimates and theirs, as we recovered all five values for the slope parameter $b$ (second column in Table 1), but only two out of five estimates of $A$ (third column in Table 1). We can only attribute these differences to typos in their manuscript, since the remaining estimates coincide and we used the same methodology.

### 3.4.2 Graphical display

We have also attempted to reproduce Figures 1 to 5 in Gavrilova et al. (2017), five graphical representations of the estimated central death rates in a logarithmic scale for each of the five subgroups in Table 1. If we focus on the estimated Gompertz parameters for the 1884-1894 cohorts born in the USA (second row in Table 1), the following values are given: 0.204 for the slope, and $9.76 \times 10^{-11}$ for the (exponential of the) intercept. We would expect these estimates to match the regression line of Figure 2 in Gavrilova et al. (2017), but this is not the case. Clearly, the slope of that line is not 0.204, and the intercept is far from $9.76 \times 10^{-11}$ (a value close to 0), which supports our claim that by "intercept parameter" they meant parameter $A$ in (2). Through a process of trial and error, we found that most of the plots displayed by Gavrilova et al. (2017) are in logarithm base 10 rather than in the natural logarithm, which is not mentioned in the manuscript. The Gompertz model becomes a linear equation when applying both the natural logarithm and the logarithm in a different base. But in this second case, a reparameterization is required, which is why the natural logarithm is used more often.

Figure 1 below reproduces Figure 2 in Gavrilova et al. (2017) by plotting the estimated central death rates in logarithm base 10. The data comprise 145 individuals born in the USA in 1884-1894 who died between the ages of 110 and 115 (IDL 2017). The central death rates, as well as the data used for their estimation, are shown in Table 2.

**Table 2.** Central death rates estimated using (1). Supercentenarians from birth cohorts 1884-1894 born in the USA (Source: IDL 2017).

| $x$ | 110 | 111 | 112 | 113 | 114 | 115 |
|---|---|---|---|---|---|---|
| $N_x$ | 145 | 76 | 41 | 19 | 5 | 1 |
| $D_x$ | 69 | 35 | 22 | 14 | 4 | 1 |
| $q_x$ | 0.4759 | 0.4605 | 0.5366 | 0.7368 | 0.8000 | 1.000 |
| $m_x$ | 0.6244 | 0.5983 | 0.7333 | 1.1667 | 1.3333 | 2.0000 |
| $\log_{10}(m_x)$ | −0.2045 | −0.2231 | −0.1347 | 0.0669 | 0.1249 | 0.3010 |

Line $y_1$ in Figure 1 corresponds to the Gompertz parameters estimated by Gavrilova et al. (2017) as shown in Table 1, and using the fact that $\ln(9.76 \times 10^{-11}) = -23.05$. Line $y_3$ are these same parameters transformed into logarithm base 10, whereas $y_2$ is a simple linear regression among the estimated central death rates in logarithm base 10. Surprisingly, we find that the line that best reproduces the original figure is $y_2$, which suggests that Gavrilova and colleagues first estimated the Gompertz parameters using a weighted non-linear regression ($b = 0.204, A = 9.76 \times 10^{-11}$); then transformed the estimated rates into logarithm base 10; and, finally, plotted the linear regression among these transformed rates while ignoring the estimated Gompertz parameters. Unfortunately, Gavrilova and colleagues do not provide any values for their estimated rates, or the equations of the regression lines of the plots. Thus, the reader's analysis is limited to a visual inspection of their graphs. But if we are right in our assessment, which seems to be the most reasonable explanation, the approach they used to produce their plots is unorthodox and misleading.

**Figure 1.** Estimated central death rates in logarithm base 10 from Table 2 (as in Fig. 2 in Gavrilova et al. 2017). Supercentenarians from birth cohorts 1884-1894 born in the USA. Line $y_1$ corresponds to the Gompertz parameters estimated by Gavrilova et al. (2017) with a weighted non-linear regression (Table 1). Line $y_3$ are these same parameters transformed into logarithm base 10, whereas $y_2$ is a simple linear regression among the estimated central death rates in logarithm base 10, and is the line that best reproduces the original graph by Gavrilova and colleagues (Source: IDL 2017).

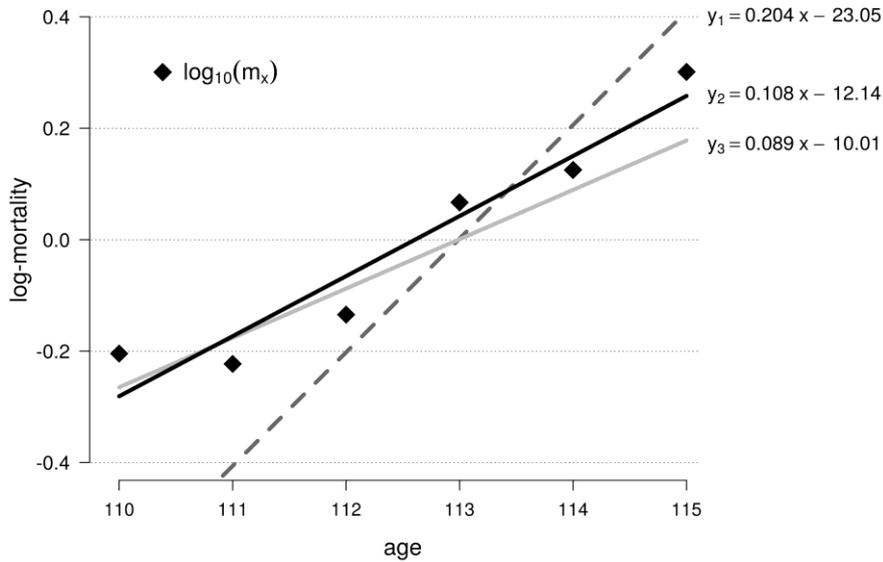

We have also been able to reproduce Figures 1, 3, 4, and 5 from Gavrilova et al. (2017) following the same procedure: first, we estimate the central death rates using (1); second, we plot the corresponding values in logarithm base 10; and, finally, we add the regression line among those transformed rates (the reader is referred to the R code in the supplementary materials to generate these figures). Furthermore, in their Figures 1 to 3, 5, and 9, the central death rate of the highest age is always around 0.3. Not by accident, $\log_{10}(2) = 0.301$, which confirms our suspicion that they assumed the probability of death at age 115 to be 1 (except in Figure 5, in which the highest age is 122), and estimated the corresponding central death rate at 2. It should also be noted that in their Figures 4, 7, and 8, the death rate for the oldest age in logarithm base 10 is 0.903, which corresponds to a central death rate of 8. This astronomically high rate is an artifact of applying (1) with an interval of length $\Delta x = 0.25$ to account for the quarter-year age scale. The death rates of 8 for the last age groups in their Figure 12 confirm this interpretation.

### 3.4.3 Confidence intervals of the central death rates

Figure 2 below displays central death rate estimates for the 1884-1894 birth cohorts on a quarter-year age scale (as in Fig. 4 in Gavrilova et al. 2017). The vertical lines represent 95% confidence intervals, and illustrate how uncertain the point estimates of the central death rates for the oldest old are given the scarcity of data. Regrettably, Gavrilova and colleagues did not take this uncertainty into account when estimating the Gompertz parameters. On the contrary, they highlight that "[i]t is also interesting to note that at very old ages (114 to 115 years), hazard rates grow in fact more steeply than predicted by the Gompertz law" (Gavrilova et al. 2017, p. 6).

**Figure 2.** Central death rates estimated using (1) in logarithm base 10, measured on a quarter-year age scale (as in Fig. 4 in Gavrilova et al. 2017). Supercentenarians from the 1884-1894 birth cohorts. Vertical lines represent 95% empirical confidence intervals obtained from data simulation. Note that the point at age 115 has a value of $\log_{10}(8) = 0.9031$. For $x = 114.75$ the confidence interval is $(-\infty, 0.9031]$, although the lower bound is not displayed in the graph (Source: IDL 2017).

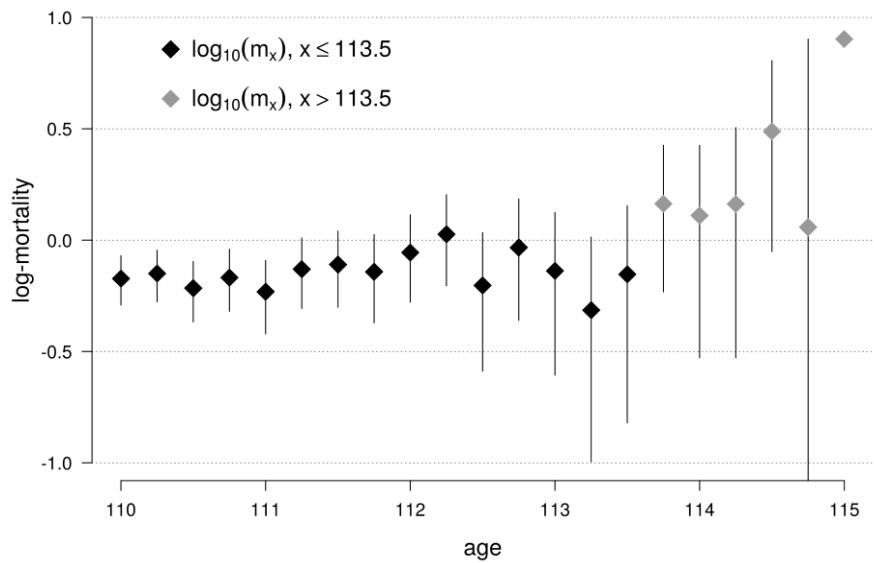

Due to the low number of observations at the oldest ages, the assumptions of the central limit theorem do not hold, and the standard methods used to compute confidence intervals are not valid. Hence, we carried out data simulation to extract empirical confidence intervals (additional details on the simulation process are provided in the Appendix, and the R code is available in the supplementary materials). For instance, we observe that the 95% empirical confidence interval for the probability of death at 114.75 is [0, 1]. Only one death was registered in this age category for the 1884-1894 birth cohorts (IDL 2017), which increases the level of uncertainty to the point where any probability of death is plausible. Following (1), the corresponding empirical confidence interval for the central death rate is [0, 8] (quarter-year age scale, $\Delta x = 0.25$), which leads to a 95% empirical confidence interval of $(-\infty, 0.9031]$ for the central death rate in logarithm base 10. On the other side of the spectrum, when it is (wrongly) assumed that the probability of death at age 115 is 1, there is no uncertainty about the death rate. For this reason, no confidence interval is shown in Figure 2 for that age (see also Table 3 in the Appendix).

Figure 2 also distinguishes between estimates for ages up to 113.5 (black dots), and estimates for ages above 113.5 (grey dots). We are aware that this division is arbitrary, but we wanted to test whether excluding the death rates for the oldest ages could affect the Gompertz estimates, since only 26 individuals from the 1884-1894 birth cohorts lived beyond age 113.5 (IDL 2017). Furthermore, the estimates of the slope parameters are much too high for these cohorts (see Table 1), with values around 0.2 that allow for a steep increase in the force of mortality after age 110, and that are also driven by the (wrong) assumption that the probability of death at age 115 is 1. By fitting the Gompertz model to the central death rates for ages 110 to 113.5 with a weighted non-linear regression, we obtain a slope parameter $b = 0.063$, a value that is notably lower than the one obtained by Gavrilova et al. (2017) using the whole range of ages ($b = 0.214$, fourth row in Table 1). Most importantly, our estimate is not statistically significant, and has a $p$-value of 0.139. Thus, we cannot reject the hypothesis that mortality is flat between ages 110 and 113.5, even when using the same methodology but excluding the oldest ages for being too sparse.

## 3.5 Conclusions

The results obtained by Gavrilova and colleagues in the analysis of data from the International Database on Longevity (IDL 2017) can be summarized as follows. Focusing on the 1884-1894 birth cohorts, from which no individuals survived beyond 115, they assigned a probability of death $q_x = 1$ to that age (Figs. 1 to 4 in their manuscript). Their corresponding estimates for the slope parameter of the Gompertz model were significantly different from 0 (see Table 1), which led them to claim that "hazard rate estimates […] continue to grow after age 110 years and follow the Gompertz law" (Gavrilova et al. 2017, p. 14). For older cohorts born before 1885, they assigned $q_x = 1$ to 122 – the highest observed age – instead of 115 (Fig. 5 in their manuscript), obtaining a slope parameter that is not significantly different from 0 (last row in Table 1), and concluding that the hypothesis of flat mortality could not be rejected in this case.

We argue that these results are a consequence of the data selection, since the division of the supercentenarians into two groups is arbitrary. This is especially relevant because, in combination with their simplistic statistical approach, it implies a change in the age to which the probability of death $q_x = 1$ is assigned. Moreover, Gavrilova et al. (2017) do not provide any measure of uncertainty for the estimated central death rates – which is imperative given the low number of observations at oldest ages – and the results and the plots they present are confusing and misleading. Overall, their work has too many inaccuracies for us to consider their conclusions reliable. Having been able to reproduce some of their graphs does not imply that the methodology used is adequate. We believe that a maximum likelihood approach including censored and truncated observations would have been more appropriate for estimating death rates.

In view of the above, does the risk of death continue to rise after age 110? The only conclusion we can reach for now is that Gavrilova and colleagues have not proved that to be the case. We hope that future research on the updated data from the IDL that is about to be released (Gampe 2018, and Jdanov et al. 2018 in this volume) will shed light on the mortality trajectories of supercentenarians.

## Acknowledgments

We are grateful to James Vaupel for encouraging our research and for providing extensive feedback. Thanks also to Jonas Schöley and an anonymous reviewer for their constructive comments, and to Miriam Hills for proofreading the manuscript.

## Appendix

This appendix describes the procedure designed to carry out the data simulation and compute the 95% empirical confidence intervals of the central death rates in logarithm base 10 displayed in Figure 2. The basic idea is to simulate 10,000 times the lifetimes between ages 110 and 115 of 400 individuals who die according to some age-specific theoretical probabilities of death. By recording the observed empirical probabilities across all simulations, we are able to compute empirical confidence intervals for each age category.

Using data on supercentenarians from the 1884-1894 birth cohorts (IDL 2017), we obtain a set of theoretical probabilities of death $q_x = D_x/N_x$ for each age category, measured in a quarter-year age scale. $D_x$ denotes the number of deaths, whereas $N_x$ is the number of exposures within each age category (see Table 3 below). Next, we create a population of 400 individuals (the same size as the abovementioned cohorts from the IDL) who are exposed to these theoretical probabilities between ages 110 and 115. In each age category, we assign to all living individuals a random number between 0 and 1 drawn from a uniform distribution: those who get a value below the corresponding theoretical probability of death die; otherwise, they live and move to the next age category. This allows for the calculation of an empirical probability of death for each age category, depending on the number of deaths and exposures observed in each case. Due to randomness, these empirical probabilities are likely to differ from the theoretical probabilities in all age categories except for the last one, since the probability of death at age 115 is set at 1.

This procedure is repeated 10,000 times, obtaining 10,000 estimates of the probability of death for each age category. We then compute the corresponding death rates following (1), and transform them into logarithm base 10. Out of this set of estimates, we compute the 95% empirical confidence intervals of the death rates in logarithm base 10 for each age category. The results are shown in Table 3, and the R code to reproduce the data simulation is available in the supplementary materials.

**Table 3.** Central death rates estimated using (1) in logarithm base 10, measured in a quarter-year age scale. Empirical 95% confidence intervals are obtained from data simulation. Supercentenarians from the 1884-1894 birth cohorts (Source: IDL 2017).

| $x$ | $N_x$ | $D_x$ | $q_x$ | $m_x$ | $\log_{10}(m_x)$ | 95% CI |
|---|---|---|---|---|---|---|
| 110 | 400 | 62 | 0.1550 | 0.6721 | −0.1726 | [−0.2919, −0.0696] |
| 110.25 | 338 | 55 | 0.1627 | 0.7085 | −0.1496 | [−0.2772, −0.0453] |
| 110.5 | 283 | 40 | 0.1413 | 0.6084 | −0.2158 | [−0.3674, −0.0952] |
| 110.75 | 243 | 38 | 0.1564 | 0.6786 | −0.1684 | [−0.3194, −0.0407] |
| 111 | 205 | 28 | 0.1366 | 0.5864 | −0.2318 | [−0.4212, −0.0913] |
| 111.25 | 177 | 30 | 0.1695 | 0.7407 | −0.1303 | [−0.3075, 0.0099] |
| 111.5 | 147 | 26 | 0.1769 | 0.7761 | −0.1101 | [−0.3010, 0.0411] |
| 111.75 | 121 | 20 | 0.1653 | 0.7207 | −0.1422 | [−0.3715, 0.0250] |
| 112 | 101 | 20 | 0.1980 | 0.8791 | −0.0560 | [−0.2782, 0.1139] |
| 112.25 | 81 | 19 | 0.2346 | 1.0629 | 0.0265 | [−0.2052, 0.2041] |
| 112.5 | 62 | 9 | 0.1452 | 0.6261 | −0.2034 | [−0.5883, 0.0339] |
| 112.75 | 53 | 11 | 0.2075 | 0.9263 | −0.0332 | [−0.3602, 0.1852] |
| 113 | 42 | 7 | 0.1667 | 0.7273 | −0.1383 | [−0.6066, 0.1249] |
| 113.25 | 35 | 4 | 0.1143 | 0.4848 | −0.3144 | [−0.9945, 0.0139] |
| 113.5 | 31 | 5 | 0.1613 | 0.7018 | −0.1538 | [−0.8212, 0.1549] |
| 113.75 | 26 | 8 | 0.3077 | 1.4545 | 0.1627 | [−0.2326, 0.4260] |
| 114 | 18 | 5 | 0.2778 | 1.2903 | 0.1107 | [−0.5283, 0.4260] |
| 114.25 | 13 | 4 | 0.3077 | 1.4545 | 0.1627 | [−0.5283, 0.5051] |
| 114.5 | 9 | 5 | 0.5556 | 3.0769 | 0.4881 | [−0.0512, 0.8062] |
| 114.75 | 4 | 1 | 0.2500 | 1.1429 | 0.0580 | (−∞, 0.9031] |
| 115 | 3 | 3 | 1.0000 | 8.0000 | 0.9031 | - |